\documentclass[sigconf]{acmart}
\usepackage{todonotes}
\usepackage{enumitem,varwidth, hyperref}
\usepackage{booktabs}

\newcommand{\change}[1]{{#1}}

\usepackage{array}
\newcolumntype{C}[1]{>{\centering\let\newline\\\arraybackslash\hspace{0pt}}m{#1}}

\AtBeginDocument{%
  \providecommand\BibTeX{{%
    \normalfont B\kern-0.5em{\scshape i\kern-0.25em b}\kern-0.8em\TeX}}}

\copyrightyear{2024}
\acmYear{2024}
\setcopyright{rightsretained}
\acmConference[CHI '24]{Proceedings of the CHI Conference on Human Factors in Computing Systems}{May 11--16, 2024}{Honolulu, HI, USA}
\acmBooktitle{Proceedings of the CHI Conference on Human Factors in Computing Systems (CHI '24), May 11--16, 2024, Honolulu, HI, USA}
\acmDOI{10.1145/3613904.3642878}
\acmISBN{979-8-4007-0330-0/24/05}

\begin{document}

\title[GazePrompt]{GazePrompt: Enhancing Low Vision People's Reading Experience with Gaze-Aware Augmentations}


\author{Ru Wang}
\affiliation{%
  \institution{University of Wisconsin-Madison}
  \city{Madison}
  \country{USA}}
\email{ru.wang@wisc.edu}

\author{Zach Potter}
\affiliation{%
  \institution{University of Wisconsin-Madison}
  \city{Madison}
  \country{USA}}
\email{zmpotter@wisc.edu}

\author{Yun Ho}
\affiliation{%
  \institution{University of Wisconsin-Madison}
  \city{Madison}
  \country{USA}}
\email{yunho7464@gmail.com}

\author{Daniel Killough}
\affiliation{%
  \institution{University of Wisconsin-Madison}
  \city{Madison}
  \country{USA}}
\email{dkillough@wisc.edu}

\author{Linxiu Zeng}
\affiliation{%
  \institution{University of Wisconsin-Madison}
  \city{Madison}
  \country{USA}}
\email{lzeng37@wisc.edu}

\author{Sanbrita Mondal}
\affiliation{%
  \institution{University of Wisconsin-Madison}
  \city{Madison}
  \country{USA}}
\email{smondal4@wisc.edu}

\author{Yuhang Zhao}
\affiliation{%
  \institution{University of Wisconsin-Madison}
  \city{Madison}
  \country{USA}}
\email{yuhang.zhao@cs.wisc.edu}

\renewcommand{\shortauthors}{}

\begin{abstract}
Reading is a challenging task for low vision people. 
While conventional low vision aids (e.g., magnification) offer certain support, they cannot fully address the difficulties faced by low vision users, such as locating the next line and distinguishing similar words. 
To fill this gap, we present \textit{GazePrompt}, a gaze-aware reading aid that provides timely and targeted visual and audio augmentations based on users' gaze behaviors. GazePrompt includes two key features: (1) a Line-Switching support that highlights the line a reader intends to read; and (2) a Difficult-Word support that magnifies or reads aloud a word that the reader hesitates with.  
Through a study with 13 low vision participants who performed well-controlled reading-aloud tasks with and without GazePrompt, we found that GazePrompt significantly reduced participants' line switching time, reduced word recognition errors, and improved their subjective reading experiences. A follow-up silent-reading study showed that GazePrompt can enhance users' concentration and perceived comprehension of the reading contents. We further derive design considerations for future gaze-based low vision aids.


\end{abstract}



\begin{teaserfigure}
  \includegraphics[width=\textwidth]{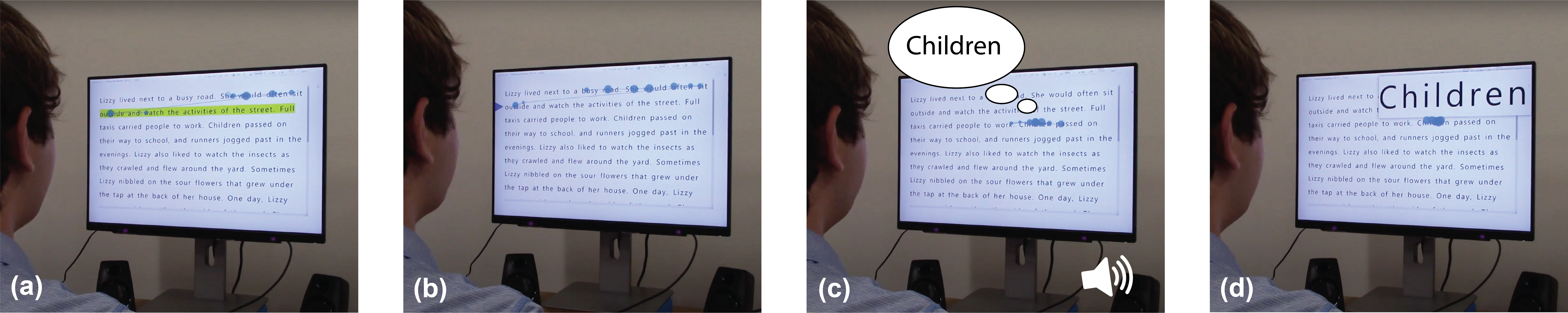}
  \vspace{-5ex}
  \caption{GazePrompt provides two key features: Line-Switching (LS) support and Difficult-Word (DW) support. Users have two design alternatives for each feature: (a) Line Highlighting and (b) Arrow for LS support, and (c) Text-to-Speech and (d) Word Magnifier for DW support. \change{The gaze visualizations are for illustration purposes only; they are not shown to the users.}}
  \label{fig:teaser}
  \Description{Figure 1: "A four-paneled teaser image of the system, GazePrompt. Each is labelled with a letter from A through D from left to right. Each panel shows a man sitting in front of a computer monitor that demonstrates a preview of a particular design feature of the system. The computer monitor shows the same text of a short story. Each panel features blue circles connected by lines, labelled gaze visualizations, to indicate the man's eye movement. The leftmost panel, labelled A, shows the second line highlighted lime green. The next panel, labelled "B" shows the same passage with a blue arrow pointing to the start of the second line. The third panel, labelled "C", shows repeated gaze behavior over the word "Children", alongside a thought bubble reading "Children". A speaker icon is shown on the bottom right of the panel. The rightmost panel, labelled "D", also shows gaze behavior lingering on the word "Children", and the word is magnified very large above the original line."}
\end{teaserfigure}

\maketitle

\section{Introduction}
Low vision is a visual impairment that cannot be fully corrected by eye glasses or contact lenses \cite{nei}. It includes a wide range of low vision conditions such as blurry vision, central vision loss, and peripheral vision loss, caused by cataract, macular degeneration, glaucoma and many more diseases. According to the WHO, at least 2.2 billion people have a near or distance vision impairment worldwide \cite{who_overview}, and that number is projected to be doubled in the next ten years \cite{who_project}. 

Reading, as a major means to access information in everyday life, can be significantly affected by different low vision conditions. For example, people with blurry vision cannot distinguish small text, while people with tunnel vision (i.e., severe peripheral vision loss) have to scan letter by letter to perceive a single word. Different low vision aids have been designed to assist low vision people, including optical and video magnifiers for print reading \cite{virgili2018reading, videomag} and built-in accessibility support on computers and smartphones for on-screen reading, such as screen magnification \cite{ioszoom, winmag} and contrast enhancement \cite{holton2014review}. These vision enhancements have also been incorporated into head-mounted displays (HMDs) to augment low vision people's residual vision in various visual tasks \cite{eSight,zhao2015foresee}. 

While conventional aids enable low vision users to read \cite{lorenzini2017impact, nguyen2009improvement}, they can also cause new barriers \cite{wang2023understanding, hallett2015reading, bruggeman2002psychophysics, moreno2021exploratory, rumney1994low}. For example, with screen magnification, low vision people still need more time to recognize words due to reduced visual span \cite{legge2001psychophysics, rumney1994low}. Visual field loss may also lead people to misidentify words as letters appear missing or distorted \cite{wang2023understanding}. Moreover, screen magnification reduces a user's field of view, thereby increasing the difficulty of locating the next line while reading long passages \cite{bruggeman2002psychophysics, moreno2021exploratory, wang2023understanding}. 


To address these issues, we seek to improve low vision people's reading performance and experience using eye-tracking technology. 
Eye tracking can be a promising solution due to its capability to recognize readers' low-level gaze behaviors and provide prompt assistance. Compared to conventional vision enhancements (e.g. magnification) that modify a user's full visual field, eye tracking presents the opportunity to render more targeted augmentations that are tailored to user behavior. 
Prior research has demonstrated early success to calibrate and collect high quality gaze data from low vision users using commercial eye trackers \cite{wang2023understanding}.
With the advance of eye tracking technology, it is critical and timely to explore how to leverage this technology and design gaze-aware augmentations for low vision people. 

We present \textit{\textbf{GazePrompt}}, a gaze-aware system that provides visual and audio augmentations based on users' gaze behaviors to support low vision people in reading tasks. Inspired by reading difficulties faced by low vision people \cite{wang2023understanding, hallett2015reading, ahn1995psychophysics}, GazePrompt focuses on two features: (1) \textit{Line-Switching Support} that augments the line a user intends to read; two design alternatives are provided due to low vision users' different visual abilities and preference, including \textit{Line Highlighting} along the whole line, and \textit{Arrow} that points out the beginning of the line; as well as (2) \textit{Difficult-Word Support} that augments the word where a user stares or hesitates around for a long time; two design options are offered, including \textit{Text-to-Speech} that reads out loud the difficult word, and \textit{Word Magnifier} that magnifies the difficult word. The features were iterated on and refined via a formative study with three low vision users. 

We evaluated GazePrompt with two studies: \change{a well-controlled reading-aloud study with 13 low vision participants to evaluate the system effectiveness quantitatively and qualitatively}, and \change{a free-form silent-reading study with another 13 low vision participants for deeper qualitative understanding of the impact of GazePrompt in a more realistic reading context (there are some participant overlap between the two studies)}. We seek to answer: (RQ1) Whether and how can GazePrompt improve low vision users' reading performance and behaviors \change{(e.g., line switching time, line switching accuracy, number of misread words)}? (RQ2) Whether and how can GazePrompt affect low vision users' subjective reading experience? (RQ3) What are low vision users' preferences on the augmentation design for each feature? 

Our research shows that \change{GazePrompt significantly reduced participants' line switching time and enabled more page scrolling flexibility in the reading-aloud study. While no significant difference, GazePrompt reduced the total number of misread words across all low vision participants. The silent-reading study highlighted that GazePrompt can enhance users' concentration and perceived comprehension. Our research also revealed low vision users' different preferences on the augmentation design and derived design considerations for future gaze-aware low vision aids}. 

\section{Related Work}




\subsection{Low Vision and Low Vision Aids}
``Low vision'' is a visual impairment that cannot be fully corrected by eye glasses or contact lenses \cite{nei}. Low vision people experience a wide range of visual impairments, such as low visual acuity, visual field loss, low contrast sensitivity loss, and extreme light sensitivity \cite{leat1999low}, which lead to various visual challenges in daily activities, such as reading \cite{legge1985psychophysics,zhao2015foresee,szpiro2016people}, navigating \cite{zhao2018looks,szpiro2016finding}, and socializing \cite{rees2007self, naraine2011social}. 

Various assistive tools and technologies are devised to support low vision people in daily tasks.
Magnifier is a cornerstone that supports people with low visual acuity, ranging from low tech optical aids (e.g., handheld magnifier and reading glasses) \cite{virgili2018reading} to electronic devices, such as video magnifiers that can magnify text or objects captured by a camera on a digital display \cite{virgili2018reading, videomag}. Prior work has also integrated magnification into a head-mounted display (HMD) to magnify real-world environments \cite{stearns2018design, zhao2015foresee}. Many personal devices (e.g., computers, smartphones) today also provide screen magnification as a standard accessibility feature \cite{ioszoom, winmag}. 

Despite its usefulness, magnification cannot address all reading barriers---low vision people still spend more time recognizing words due to reduced visual span \cite{legge2001psychophysics, rumney1994low}, and people with visual field loss still face difficulty distinguishing similar words or identifying long words due to missing or distorted letters \cite{wang2023understanding}.
Moreover, magnification itself can bring new challenges \cite{cheong2007relationship, cheong2008relationship, hallett2015reading, ahn1995psychophysics, szpiro2016people, wang2023understanding}. For example, in a reading scenario, the decreased field of view reduces a user's reading speed \cite{cheong2007relationship, cheong2008relationship}. The user also has to pan around with a mouse to reveal different proportion of a page, making it difficult to locate the next line quickly and accurately \cite{ahn1995psychophysics, wang2023understanding} as well as increasing their cognitive load \cite{szpiro2016people}. 
Researchers have come up with solutions to improve user experiences with screen magnification \cite{bigham2014accessibility, taras2010improving, aydin2020towards}; for example, Aydin et al. \cite{aydin2020towards} developed an intelligent screen magnifier that optimally magnifies elements of interest in dynamic content, as identified by a video saliency model.

Besides magnification, other types of visual augmentations are designed to compensate different visual impairments \cite{aydindougan2021applications, pamparuau2021flexisee, zhao2019seeingvr, zhao2015foresee, huang2019augmented}. For example, contrast enhancement is a common strategy to tackle low contrast sensitivity \cite{zhao2015foresee, choudhury2010color, stearns2018design}. Modern video magnifiers (e.g., RUBY \cite{videomag}) provide high-contrast color filters to enhance low vision users' visual experience. Researchers have also explored contour enhancement by increasing contrast between objects and their background, which has been effective for people with central vision loss \cite{hwang2014augmented, huang2019augmented, zhao2019seeingvr} 
A minified contour of a wider field is also used as an overlay (a.k.a., multiplexing) in one's central vision to expand the visual field of people with peripheral vision loss \cite{peli2006tunnelvision, zhao2019seeingvr, peli2001vision}. Moreover, color enhancement (e.g., changing certain color) has been designed to improve the color discerning ability for people who are color blind \cite{tanuwidjaja2014chroma, langlotz2018chromaglasses}.

In addition to image-processing-based enhancements that alter users' full field of view, researchers have started exploring more targeted visual cues based on specific contexts or tasks, such as navigation \cite{huang2019augmented, zhao2020effectiveness, zhao2019designing, fox2023using} and visual search \cite{zhao2016cuesee,lang2021pressing}. For example, Zhao et al. \cite{zhao2016cuesee} facilitated visual search tasks by rendering visual cues directly on the search targets. Fox et al. \cite{fox2023using} has explored the usability of different visual cues that highlight obstacles for low vision users in navigation. However, there has been limited research on more tailored, context-aware visual augmentations for vital but challenging reading tasks. To our knowledge, the only relevant work is Gowases et al. \cite{gowases2011text}, which presents line highlighting or a pointer to label the next line, thus mitigating the loss of context caused by magnification. However, the feature's manual control method, using a mouse for magnification manipulation and keyboard for highlight control, caused operation difficulty and increased cognitive load. Moreover, the feature has only been evaluated with sighted people, without involving any low vision users. 

As opposed to the manually-controlled visual support methods of prior work \cite{gowases2011text}, we seek to leverage eye tracking technology to  provide more intelligent, gaze-aware reading support tailored to low vision users' behaviors and intent, thus compensating the drawbacks of existing low vision aids. 

\subsection{Gaze-Aware Technology}
Eye tracking is a promising technique to enhance users' reading experience. Research efforts have been made on eye-tracking-based reading support. One important reading task is to resume from the previous reading position when switching attention between reading and other activities. Eye tracking-based solutions are thus developed to track and label the previous reading position \cite{switchback, kern2010placeholders, RSVPGo}. For example, Mariakakis et al. \cite{switchback} used eye tracking to detect when the user looks away from the phone and looks back, and highlighted the line where they have left off to direct the user's attention. Besides locating lines, eye tracking is also used to enhance comprehension \cite{hyrskykari2000idict, guo2019reading, cheng2015gazeannotation}; for instance, Hyrskykari et al. \cite{hyrskykari2000idict} generated real-time text translation for foreign readers when difficult words are detected from readers' gaze patterns. Cheng et al. \cite{cheng2015gazeannotation} has designed a gaze-based reading annotation system that summarizes and shares a teacher's reading characteristics, such as reading speed, transitions between sections, and frequency of re-reading, to improve their students' reading comprehension. Moreover, other than inferring readers' behaviors or intent, researchers also use gaze as direct control to replace traditional input methods, such as mouse and keyboard\cite{chen2019visual, lee2022vrdoc, shakil2019codegazer}. For example, Shakil et al. \cite{shakil2019codegazer} has developed a system that allows a user to use gaze to control code navigation, such as performing ``Go to Definition'' by dwelling on a color square on the side of the screen, and found that gaze control can improve code reading efficiency.

Eye tracking-based augmentations can also benefit the reading performance for people with reading disabilities \cite{lunte2020towards, sibert2000auditory}. For example, Sibert et al. \cite{sibert2000auditory} proposed a gaze-based auditory support that can highlight words and pronounce them if the user pauses at a word for a relatively longer duration. Lunte and Boll \cite{lunte2020towards} designed a gaze-contingent reading assistant for children with reading difficulties that  dynamically changes the color of letters according to the user's gaze position to improve their reading experience.

Besides reading, gaze-aware technology has also been designed to support other activities, such as collaborative work \cite{wang2019gaze, schneider2017real, vertegaal1999gaze}, social activities \cite{muller2018robust}, and video conferencing \cite{samrose2018coco, vertegaal2002gaze, he2021gazechat}. For example, He et al. \cite{he2021gazechat} developed a virtual conferencing system that conveys users' eye gaze direction to other people in a meeting through their profile picture, improving the engagement of meeting participants.

Although prior research has widely used eye tracking to enhance reading comprehension and efficiency over a variety of tasks, they mainly focus on sighted people. Little research has investigated how this technology can be applied to assist low vision people in reading tasks. 

\subsection{Eye Tracking Research for Low Vision}
Despite the potential, eye tracking research remains nascent in the low vision field. Compared to sighted people, low vision people may have different visual abilities, eye characteristics, and eye behaviors, which leads to low gaze estimation accuracy and high data loss in eye tracking \cite{maus2020gaze, wang2023understanding, manduchi2022gaze, murai2010eye}. As a result, eye tracking technology has been mostly used for vision science and ophthalmology to simulate low vision conditions for sighted participants, collecting early empirical data from participants with ``simulated low vision'' \cite{aguilar2017evaluation, gupta2018cfl, harvey2014scotoma, aguilar2012augmentedvision}.
The most commonly simulated condition is central scotoma (i.e., blind spots in one's visual field). By tracking a user's gaze, an artificial blind spot is rendered at the position that they are looking at on a computer or HMD. For instance, to evaluate the effectiveness of a text-remapping aid for people with central vision loss (a method that re-renders blocked text around the scotoma), Gupta et al. \cite{gupta2018cfl} recruited 35 sighted participants who experience simulated gaze-contingent scotoma, finding that participants with simulated scotoma read significantly faster with the remapping aid. Similarly, Aguilar and Castet \cite{aguilar2017evaluation} assessed a gaze-controlled magnifier by simulating gaze-contingent scotoma for 10 sighted participants. 
While gaze-contingent low vision simulation allows easy data collection to form preliminary test results, they are not guaranteed to transfer to low vision people due to the different viewing strategies between the two groups. For example, a person with central scotoma may have developed a preferred retinal locus (PRL) to replace the damaged fovea \cite{prahalad2020asymmetries}, which cannot be reflected by participants with simulated scotoma. 

In the field of Human-Computer Interaction (HCI), less research has investigated or leveraged eye tracking technology for low vision people. Maus et al. \cite{maus2020gaze} has designed and evaluated a gaze-controlled screen magnifier with seven low vision participants and found five of them demonstrate high data loss (> 50\%). Meanwhile, Ivanov et al. \cite{ivanov2016eye} attempted to study the gaze behaviors of people with peripheral vision loss in walking tasks via eye tracking, but 11 out of 25 participants failed in calibration. 
Recently, Wang et al. \cite{wang2023understanding} improved gaze calibration and data collection methods and gained high quality gaze data from low vision users that is comparable to sighted users, enabling researchers to investigate low vision people's gaze behaviors with a commercial eye tracker. They further analyzed and compared low vision and sighted participants' gaze data in reading tasks, revealing difficulties faced by low vision readers, such as tracking and locating a specific line and quickly identifying difficult words. Despite early successes in collecting low vision people's gaze data, no research has explored how to design effective gaze-aware technologies to enhance their visual experiences. Our research fills this gap by designing, implementing, and evaluating GazePrompt.  

\section{GazePrompt: System Design \& Implementation}
\label{System Design}
We designed and built GazePrompt, a gaze-aware system that generates visual and auditory augmentations based on a low vision user's gaze behaviors to facilitate reading. GazePrompt is a \textit{complement} to existing low vision aids (e.g., screen magnification, contrast enhancement) that further enhances people's reading experience. As such, we focus on addressing two key challenges that low vision people encounter in reading, even when using conventional low vision aids: line-switching, which is especially difficult under high magnification \cite{ahn1995psychophysics, wang2023understanding}, and difficult word recognition (e.g., visually similar words, long words) due to cut-off, missed, distorted letters caused by vision loss \cite{wang2023understanding, mackeben2015random}. We elaborate on our feature design for these two challenges (Section \ref{sec:design}), the system implementation (Section \ref{sec:implementation}), and our iteration with three low vision participants in a formative study (Section \ref{sec:formative}).

\begin{figure*}[h]
  \includegraphics[width=0.95\textwidth]{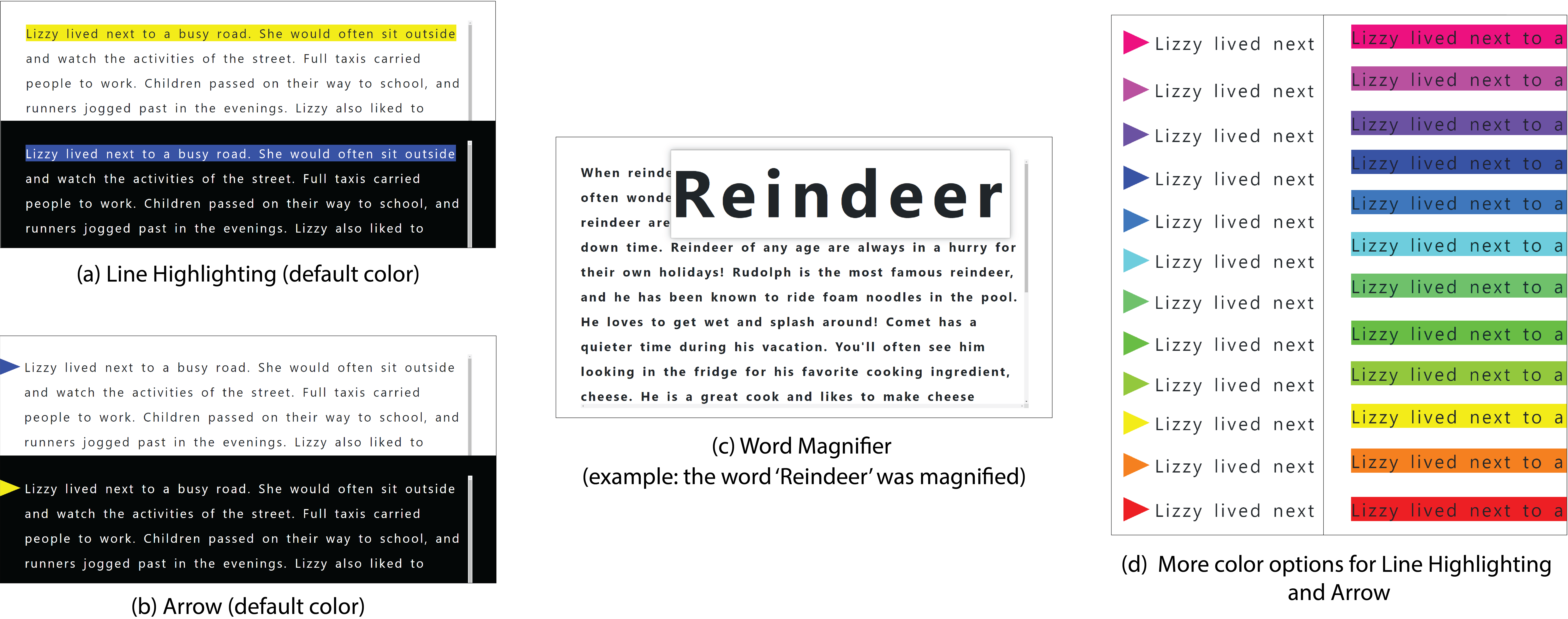}
  \caption{GazePrompt interfaces. (a) The Line Highlighting augmentation of Line-Switching Support; (b) The Arrow augmentation of Line-Switching Support; (c) The Word Magnifier of Difficult-Word Support; (d) More color options for Line-Switching Support.}
  \label{fig:design}
    \Description{Figure 2: "A figure with visualizations of the four GazePrompt interfaces labelled A through D.
Visualization A, labelled "Line Highlighting (default color), shows a short passage split in half, with black text on white background on top and white text on black background on the bottom. The first line of each passage is highlighted in yellow on top and blue on bottom, emphasizing the line "Lizzy lived next to a busy road. She would often sit outside".
Visualization B, labelled Arrow (default color), shows a short passage split in half, with black text on white background on top and white text on black background on the bottom. An arrow points to the first line of each passage, colored blue on top and yellow on bottom, emphasizing the line "Lizzy lived next to a busy road. She would often sit outside".
Visualization C, labelled Word Magnifier (example: the word 'Reindeer' was magnified), shows a different passage in bold black text on white background, with the word "Reindeer" emphasized in very large text and its own text box window above the line "Reindeer of any age are always in a hurry".
Visualization D, labelled "More color options for Line Highlighting and Arrow", shows a rainbow assortment of colors next to or overlaid onto the text "Lizzy lived next". From top to bottom, there is a pink option, a pinkish purple option, two purplish options, two blueish options, three greenish options, one yellow option, one orange option, and one red option. All of these examples are shown using black text on a white background."
}
\end{figure*}

\subsection{Feature Design} \label{sec:design}

\subsubsection{Line-Switching (LS) Support} 
To enable users to easily and accurately follow a line or locate the next line, we design a line-switching support method that \textit{detects and augments the line of interest (LOI)}--- the line that the user is reading or intends to read. Via eye tracking data, we recognize the current line the user is focusing on as well as their line switching behaviors. We distinguish three behaviors and present augmentations accordingly: (1) when a user is \textit{following a line}, the current line is augmented; (2) when the user finishes the current line and \textit{switches to the next line}, the next line is augmented right away; and (3) when the user is \textit{jumping to a different line} (e.g., skipping lines or revisiting previous lines), the target line is augmented after the line jumping behavior stabilizes. Recognition algorithms are described in Section \ref{sec:implementation}. We provide two augmentation options for different visual conditions and preferences:



\textbf{\textit{Line Highlighting.}} Prior work has shown that highlighting can improve searching and reading performance \cite{wu2003improving, switchback}, making it easier to locate a new line \cite{gowases2011text} and reducing cognitive load \cite{ozyegen2022word}. We thus highlight a LOI by changing its background color. Since low vision users usually need high contrast to read, our highlighting color is adaptive to the reading materials; by default, we use yellow (RGB [255,255,0]) to highlight black text on a white page and blue (RGB [0,0,255]) to highlight white text on a black page (Fig. \ref{fig:design}a). We further allow users to customize the highlighting colors due to feedback from the formative study (Section \ref{sec:formative}). 

\textbf{\textit{Arrow.}} Low vision users may not want the whole line to be highlighted since it reduces the contrast (black-white has the highest contrast) and may also distract the users \cite{draxler2023useful}. We thus provide a relatively subtle design--- labeling the beginning of a LOI with a high-contrast arrow--- to help a low vision user to locate the next line \cite{gowases2011text}. 
Similar to the highlighting augmentation, we use a blue arrow to create a high contrast against the white background, and a yellow arrow for black background. Users also have the flexibility to customize the color (Fig. \ref{fig:design}b).

\subsubsection{Difficult-Word (DW) Support} 
To enable a low vision user to accurately and quickly recognize a word, we design a difficult-word support that \textit{detects and augments the word of interest (WOI)}--- the word that the user is interested in but has difficulty recognizing. Based on eye tracking data, we detect a word as a WOI when the user stares or hesitates on the same word for a long time (see implementation in Section \ref{sec:implementation}). Two augmentation alternatives are provided for a WOI: 


\textbf{\textit{Word Magnifier.}} Since magnification is the most common method used by low vision people to see details, we magnify the WOI to the maximum magnification level supported by the screen magnifier to enable the user to thoroughly examine the word. The magnified version is rendered above the WOI (or below if the word is close to the upper border of the display) to avoid blocking the reading context. The magnifier will disappear if the user moves their eyes away from it. 
The word magnifier provides local magnification of the WOI while maintaining the global reading context, which could be useful for low vision users who do not prefer full-screen magnification (Fig. \ref{fig:design}c). 

\textbf{\textit{Text-to-Speech.}} Besides visual augmentations, some low vision people prefer auditory feedback on complex information since it reduces their visual effort \cite{zhao2020effectiveness}. 
We thus design an auditory augmentation that reads aloud the WOI to the user. Similar design has been applied to support reading for people with dyslexia and proven to be effective \cite{rasmussen2013fusing, sibert2000auditory}. 

\begin{figure*}[h]
  \includegraphics[width=0.95\textwidth]{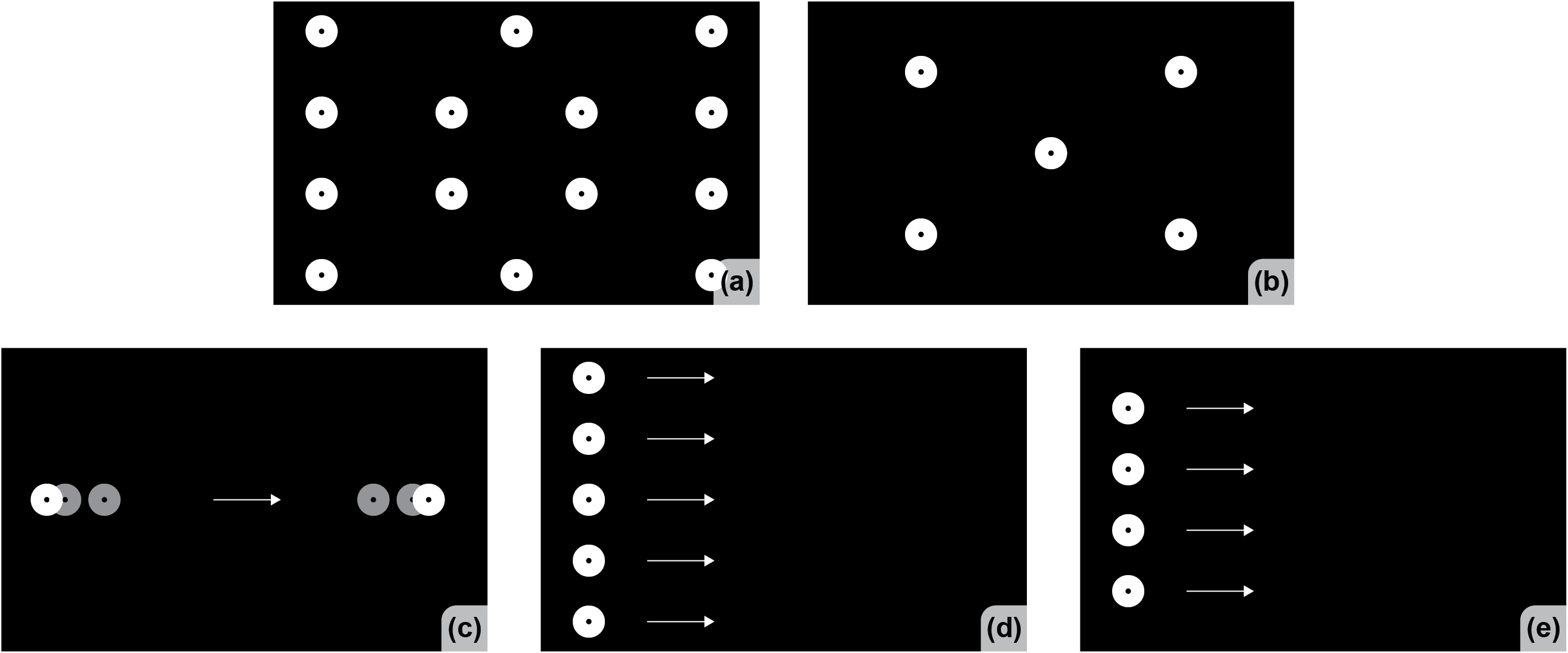}
  \caption{Calibration \& Validation interfaces. (a) 14-dot calibration interface; (b) 5-dot validation interface; (c) An illustration of sliding target for line-based calibration (the target will move from left side to the right side of the screen once activated, the white arrow is for demonstration only); (d) 5-line calibration interface; (e) 4-line validation interface.}
  \label{fig:cali}
   \Description{Figure 3: "A 5-section figure with each panel labelled A through E. Each panel shows a black rectangle with a varying amount of white circles within. Each white circle has a black dot in the center. Panel A, labelled "14-dot calibration", shows 14 circles laid out in four rows, three across on top and bottom and four across in each of the middle two rows. Panel B, labelled "5-dot validation", shows five circles laid out in an X formation, with the center dot in the center of the black background. Panel C, labelled "Illustration of sliding targets for line-calibration \& validation", demonstrates a white circle moving from the left side of the image to the right side. A few gray "afterimages" of the circle show that the circle accelerates as it starts and decelerates as it approaches the right side. Panel D, labelled "5-line calibration", shows 5 circles vertically stacked on top of one another, evenly spaced from 10\% to 90\% of the height of the rectangle. Each circle has a white arrow to the right of it pointing rightwards. Panel E, labelled "4-line calibration", shows 4 circles vertically stacked on top of one another, evenly spaced from 20\% to 80\% of the entire height of the rectangle. Each circle has a white arrow to the right of it pointing rightwards."
}
    \vspace{-2ex}
\end{figure*}

\subsection{Implementation} 
GazePrompt was implemented using a Tobii Pro Fusion (120Hz) \cite{tobiiprofusion} screen-based eye tracker attached on the bottom of a computer display (24-inch, $1920 \times 1200$ resolution). We built the system through three steps: (1) Improving the eye tracking calibration for low vision users;  (2) Filtering and processing the gaze data; and (3) Recognizing gaze behaviors. We describe the implementation of each step below.  

\subsubsection{Gaze Calibration \& Collection.} \label{sec:cal}
Low vision users experience inaccurate recognition and high data loss with eye trackers due to inaccessible calibration and data collection methods \cite{maus2020gaze,murai2010eye}. To address this issue, we adopted an adjustable calibration interface (i.e., the calibration target size was adjusted based on user's visual ability) and dominant-eye-based data collection (i.e., gaze data collection focused on the user's dominant eye if there was one) following guidance from prior literature \cite{wang2023understanding}. We used a 14-dot calibration interface, the maximum target number supported by the Tobii Pro SDK, to achieve a high calibration granularity for low vision users (Fig. \ref{fig:cali}a), followed by a 5-dot validation interface (Fig. \ref{fig:cali}b). 

Moreover, since eye tracking suffers from vertical drift--- the vertical coordinate of the estimated gaze position becomes less accurate overtime \cite{carr2022algorithms, shamy2023identifying}--- we improved the calibration process by involving a \textit{line-based correction} after the 14-dot calibration. 
Specifically, we rendered a target stimulus (a white solid circle with a black dot in the center) that moved horizontally along a line from the left side of the screen to the right. We instructed the user to keep tracking the target with their gaze until it disappeared (Fig. \ref{fig:cali}c). The target was the same size as the ones used in 14-dot calibration. 
The same process was repeated five times, with the vertical position of the calibrated line shifting down by 20\% of the screen height each time (Fig. \ref{fig:cali}d). 
We collected the user's gaze data as they tracked each line and calculated the mean vertical gaze offset from the line \change{as the vertical drift of that line}. The system then interpolated \change{the vertical drift} in between adjacent calibrated lines and \change{correct the drift} across the entire screen. A validation interface with another four \change{horizontally moving} targets were then presented to evaluate the vertical correction (Fig. \ref{fig:cali}e).

\subsubsection{Gaze Data Filtering and Processing.} The gaze data was retrieved via the Tobii Pro SDK \cite{tobiiprosdk} in Python. We filtered the data by removing noise and outliers on the fly. We then converted the user's continuous gaze data into a series of fixations (i.e., short pauses of gaze during reading to process information \cite{salvucci2000identifying}) via a real-time fixation detection algorithm \cite{kumar2008improving, feit2017toward} for further gaze behavior detection. We set up a Flask-SocketIO \cite{flask-sio} server to process the gaze data and enable bi-directional and low-latency communication between the server and the system user interface. 

\subsubsection{Gaze Behavior Recognition.} \label{sec:implementation}
We then recognized high-level gaze behaviors upon the fixation sequence. Specifically, we identified the current reading line and line switching behaviors for LS support, and we recognized hesitation around words for DW support.  

\begin{figure*}[h]
  \includegraphics[width=0.95\textwidth]{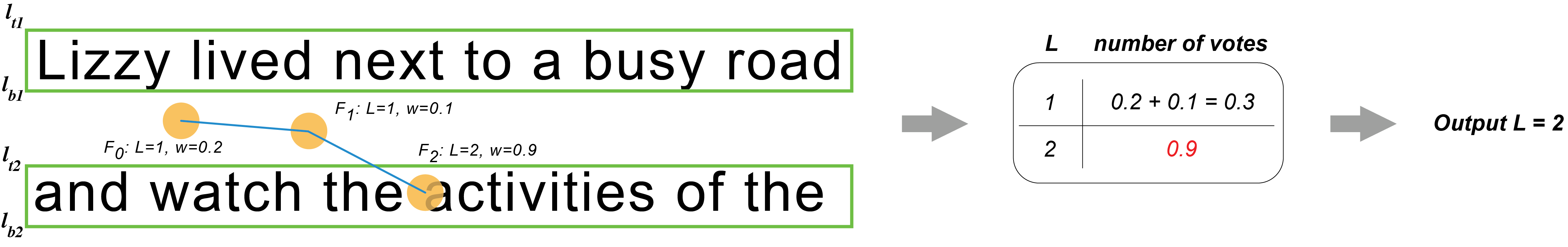}
  \caption{An example of line identification. The green bounding boxes represent the space of two lines: L1 and L2. Orange dots represent fixations and blue line segments represent saccades. The line IDs that are closest to the latest three fixations are 1, 1, and 2, with weight 0.2, 0.1, and 0.9, respectively. The final landing line is line 2 because it receives the most votes.}
  \label{fig:lid}
    \Description{Figure 4: "A figure with three sections with bold black arrows pointing to the right between them. The leftmost section consists of two lines of text, the top reading "Lizzy lived next to a busy road", and the bottom reading "and watch the activities of the". The left side of each line has text reading "L subscript T 1" towards the top of the first line, "L subscript B 1" towards the bottom of the first line, "L subscript T 2"towards the top of the second line and "L subscript B 2" towards the bottom of the line. Overlaid on the two lines of text are three orange circles joined by a blue line. The left circle is located above word "watch" and labelled F0: L = 1, w = 0.2. The middle circle is located above the first "the" and is labelled "F1: L = 1, w = 0.1". The right circle is located overlapping the "a" in word "activities" and is labelled "F2: L=2, w = 0.9". The second section of the three sections in this figure show a 2 by 2 chart with the left column labelled "L" and the right column labelled "number of votes". The first row labelled "1" contains the equation 0.2 + 0.1 = 0.3. The second row labelled "2" contains the text 0.9 highlighted in red.
The third section of this figure just consists of bold text reading "Output L = 2"."
}
\end{figure*}

\textbf{\textit{Line Identification.}} Considering the inherent uncertainty of eye tracking, we used a weighted voting mechanism \cite{jung2011improving} to identify the line that a user is reading based on their latest three fixations. We defined the space of a line with a bounding box that wrapped the line of text, so that a line $L$ can be defined by the top and bottom border positions of its bounding box $(l_t, l_b)$, as shown in Fig. \ref{fig:lid}. For each fixation $F (f_x, f_y)$, we first calculated its landing line $L (l_t, l_b)$ by identifying the line that had the smallest vertical distance to the fixation. The landing line thus represented the ``vote'' of the fixation. We then calculated the weight ($w$) of the fixation: $w = \frac{1}{1 + |d|}$, where $d$ represented the normalized distance between the fixation and its landing line; so that the smaller the normalized distance was, the more weight a fixation had. The normalized distance was defined as: $d = \frac{f_y - m}{0.5h}$, where $m = \frac{l_b + l_t}{2}$, \change{$h = l_b - l_t$}. Based on the vote and weight (i.e., number of votes) of each fixation, we determined the final landing line as the line that had the most votes from the latest three fixations.




\textbf{\textit{Line Switching Behaviors.}} We detected a line switching behavior via a return sweep--- fast eye movements to switch focus from the end of one line to the beginning of the next \cite{rayner1998eye}. Suppose $FA (fa_x, fa_y)$ and $FB (fb_x, fb_y)$ are two adjacent fixations. We determined a return sweep based on the following criteria: (1) the horizontal distance between the two fixations should be greater than a threshold $T^{LS}_0$, i.e., $fb_x - fa_x > T^{LS}_0$ ; (2) the later fixation should land on the left portion of the page, around the beginning of a line, i.e., $fb.x < T^{LS}_1$; and (3) the two fixations should be at least one line apart vertically, i.e., $fb_y - fa_y > T^{LS}_2$. Inspired by prior work on line switching detection \cite{carr2022algorithms}, we set $T^{LS}_0$ to be $500 px$, $T^{LS}_1$ to be one third of the text width, and $T^{LS}_2$ to be the line height.

When no return sweep was detected, we distinguished a line following behavior and a line jumping (i.e., jumping to a different line without return sweep) behavior. Specifically, if the line identification result remained the same, we assumed that the user was following a line; if the line identification result changed and the change remained stable for three consecutive fixations, we treated it as a line jumping behavior.

\textbf{\textit{Difficult Word Detection.}}
Fixation duration on a word is positively correlated to the difficulty of information processing, and both first fixation duration and total fixation duration on a word can indicate the difficulty of recognizing the word \cite{rayner1998eye, inhoff1984two}. Moreover, the number of re-fixations (i.e. other than the first fixation) on a word indicates the amount of adjustment towards optimal viewing location \cite{rayner1998eye}, which is also related to word recognition difficulty. Therefore, our system detected the difficulty of word recognition when any of the following happened: (1) the first fixation on a word was longer than threshold $T^{DW}_0$; (2) the number of re-fixations on the word was greater than threshold $T^{DW}_1$; (3) the total fixation duration on a word was greater than threshold $T^{DW}_2$. Note that, to facilitate real-time difficult word detection, we only considered fixations on a word in one-pass (i.e., consecutive fixations on a word before leaving for other words). We empirically determined the thresholds as: $T^{DW}_0 = 500$ ms,  $T^{DW}_1 = 4$, and $T^{DW}_2 = 1500$ ms according to the data collected in our prior work \cite{wang2023understanding}.


Building upon gaze behavior recognition, GazePrompt rendered corresponding augmentations on a web-based reading interface. We built the interface using React \cite{react}. 

\subsection{Design Iteration via a Formative Study} \label{sec:formative}
Following a user-centered design approach \cite{abras2004user}, we conducted a formative study with three low vision participants (P1-P3 in Table \ref{tab:lv_dem}) to iterate on the design of GazePrompt. We introduced GazePrompt's two features and their two design alternatives to participants.  They then freely read short passages with our system until they were familiar with each feature. The passages were magnified and adjusted to the most suitable contrast to ensure readability for participants. 
We then asked about their experience with the two features and how they wanted to improve them. We analyzed their responses qualitatively and refined our feature design accordingly:

\textbf{\textit{More Color Selection for Line-Switching Support.}} All three participants pointed out that the default color options (i.e., yellow or blue) in the LS support were not their preferred colors; for example, P2 felt that yellow was too bright, expediting eye strain. We thus improved our system by introducing more color options. To simplify color adjustment, our color selection procedure adopted the HSL (hue, saturation, and lightness) model \cite{wu2003improving}, where users can choose their preferred hue and lightness with the saturation remaining 100\% (Fig. \ref{fig:design}d).

\textbf{\textit{Adjustable Triggering Threshold for Difficult-Word Support.}} Participants preferred different triggering time for DW support due to their different visual abilities and reading habits. P3 mentioned the feature was triggered too late, while P2 felt it triggered too frequently. Therefore, we made the fixation duration thresholds $T^{DW}_0$ and $T^{DW}_2$ adjustable, 
with $500 ms$ and $1500 ms$ being the starting points, allowing users to adjust the value up or down with a step of $50 ms$ and $250 ms$, respectively. 

\begin{table*}[t]
\scriptsize
\centering
\begin{tabular}
{C{0.4cm}C{0.5cm}C{2.5cm}C{0.7cm}C{1cm}C{2cm}C{2cm}C{2.4cm}}
\toprule
\textbf{ID}&\textbf{Age/{\newline}Gender}&\textbf{Diagnosed{\newline}Condition}&\textbf{Legally {\newline} Blind} &\textbf{Visual Acuity}&\textbf{Visual Field}&\textbf{Other Visual {\newline} Difficulties}&\textbf{Accessibility {\newline} Tech Used}\\
\midrule
P1 & 28/F & Stargardt's Disease & Y & L: 20/200{\newline} R: 20/200 & Central vision loss & Sensitive to light & Large font, full-screen magnifier, eSight \\

\hline
P2 & 78/F & Retinitis Pigmentosa & Y & L: 20/25{\newline} R: 20/80 & Peripheral vision loss & Sensitive to light & Brighter and Bigger \\

\hline
P3 & 54/M & Stargardt's Disease & N & L: 20/200{\newline} R: 20/125 & Central vision loss & Sensitive to light & Large font, text-to-speech, full-screen magnifier \\

\hline
P4 & 64/F & Macular Degeneration, Cataract & N & L: 20/400{\newline} R: 20/125 & Central vision loss & Sensitive to light, color looks grayish & Large font, lens magnifier \\

\hline
P5 & 57/F & Retinitis Pigmentosa & Y & L: 20/100{\newline} R: 20/100 & Peripheral vision loss & Sensitive to light, color blind & Brighter and Bigger, large font, screen magnifier \\

\hline
P6 & 69/F & Cataract & N & L: 20/200{\newline} R: 20/80 & Intact & N/A & N/A \\

\hline
P7 & 79/F & Macular Degeneration Intermediate Dry & N & L: 20/50{\newline} R: 20/50 & Central vision loss on left eye & Sensitive to light & Large font \\

\hline
P8 & 57/M & Optic Neuropathy & Y & L: 20/160{\newline} R: 20/400 & Peripheral vision loss & Sensitive to light, difficulty with shades of red and green & Screen magnifier, Invert color, large font \\

\hline
P9 & 87/F & Macular Degeneration & N & L: 20/40{\newline} R: 20/25 & Central vision loss & Sensitive to light, difficulty with blues and greens & Large and bold font \\

\hline
P10 & 65/M & Non-arteritic anterior ischemic optic neuropathy (NAION) & N & L: 20/20{\newline} R: 20/50 & Central vision loss on right eye, peripheral vision on left eye & Sensitive to light & Large font \\

\hline
P11 & 81/F & Macular Degeneration & N & L: 20/200{\newline} R: 20/60 & Central vision loss & Sensitive to light & Large font \\

\hline
P12 & 61/F & Scar tissue on Retina & N & L: 20/125{\newline} R: 20/100 & Central vision loss & Sensitive to light & Large font \\

\hline
P13 & 30/F & Retinitis Pigmentosa & N & L: 20/50{\newline} R: 20/25 & Peripheral vision loss & N/A & Large and bold font, night-time mode, \\

\hline
P14 & 71/M & Macular Degeneration & N & L: 20/250{\newline} R: 20/200 & Central vision loss & N/A & Large font, invert color \\

\hline
P15 & 61/F & Spinal Meningitis & Y & L: 20/2200{\newline} R: 20/200 & Peripheral vision loss & Difficulty with navy blue and black & Full-screen magnifer, Large font \\

\hline
P16 & 30/F & Congenital Glaucoma, Cataract & N & L: 20/400{\newline} R: 20/100 & Peripheral vision loss & Sensitive to light & Large and bold font, invert color \\

\hline
P17 & 63/F & Cataract, Retina Scarring & Y & L: 20/80{\newline} R: 20/125 & Central vision loss on right eye & Sensitive to light & Brighter and Bigger, BARD \\

\hline
P18 & 85/F & Macular Degeneration & N & L: 20/400{\newline} R: 20/30 & Central vision loss on left eye & Sensitive to light & N/A \\

\hline
P19 & 72/F & Retinitis Pigmentosa & Y & L: 20/40{\newline} R: 20/40 & Peripheral vision loss & N/A & High brightness at night, large cursor \\

\hline
P20 & 37/M & Macular Dystrophy & N & L: 20/30{\newline} R: 20/40 & Central vision loss & N/A & Large font, high brightness \\

\hline
P21 & 34/F & Retinopathy of Prematurity, Peripheral Retinal Degeneration & Y & L: 20/400{\newline} R: 20/200 & Peripheral vision loss & Sensitive to light, difficulty with navy blue and black & Voiceover, ZoomText, JAWS, full-screen magnifier \\

\hline
P22 & 82/F & Macular Degeneration & N & L: 20/160{\newline} R: 20/200 & Central vision loss & Sensitive to light & Zoom-in \\

\hline
P23 & 61/F & Ocular Melanoma & N & L: 20/30{\newline} R: 20/50 & Peripheral vision loss on left eye & Sensitive to light & High contrast, bold and large font, low brightness, BARD, screen magnifier \\

\hline
P24 & 76/F & Glaucoma & N & L: 20/60{\newline} R: 20/60 & Peripheral vision loss & Sensitive to light & High contrast, large font \\

\hline
P25 & 60/M & Cortical Blindness & Y & L: 20/100{\newline} R: 20/200 & Peripheral vision loss & Sensitive to light, no color vision & Screen reader, Alexa, Siri \\
\bottomrule
\end{tabular}
\caption{\change{Participant demographics: P1-P3 in the formative study, P3-P15 in Study 1, and P13-P25 in Study 2.}}
\label{tab:lv_dem}
\Description{Table 1: "This table shows a list of 25 participants including an ID (listed as P1 through P25); their age (between 28 and 87); their gender (M or F); their diagnosed condition (like Macular Degeneration, Glaucoma); whether or not they're legally blind (Yes or No); their visual acuity for each eye (like Left: 20 200, Right: 20 80); information about their visual field (like central vision loss, peripheral vision loss); any other visual difficulties (like being Sensitive to light and/or having difficulty with certain colors like blues and greens); and any other accessibility technology used in their daily life (like large font, full-screen magnifier, high contrast, or use of a screen reader)."}
  \vspace{-5ex}

\end{table*}

\section{Study I: Reading Aloud}
\label{study1}
We first evaluate the effectiveness of the two features in GazePrompt (RQ1 in Intro) \change{in a reading aloud task, which helps us to better control participants' focus, reducing the confounding effect of content comprehension on their gaze behaviors \cite{wang2023understanding}.} We also explore low vision users' subjective experiences (RQ2) and customization preferences (RQ3) when using GazePrompt. To assess the feature effectiveness quantitatively, \change{we compare participants' reading behaviors between using GazePrompt and not using GazePrompt (the baseline). Since GazePrompt is developed as a complement to existing low vision aids (e.g., increasing font size) instead of replacing them, we allowed participants to change font size and use contrast enhancement in both conditions in the study.} We address two sets of hypotheses:

\begin{itemize}[leftmargin=2em]

    \item[\textbf{H1}] \textbf{The Line-Switching (LS) support can improve a low vision user's line-switching performance.}
    
    \begin{itemize}[align=left, leftmargin=0.25cm]
        \item[H1.1] Low vision people switch lines faster with LS support than the baseline.
        \item[H1.2] \change{Low vision people switch lines more accurately with LS support than the baseline.}
        
    \end{itemize}

    \item[\textbf{H2}] \textbf{The Difficult-Word (DW) support can improve a low vision user's word recognition performance.}
    \begin{itemize}[align=left, leftmargin=0.25cm]
        \item [H2.1] Low vision people's maximum time spending on a word is reduced with DW support than the baseline.
        \item [H2.2] \change{Low vision people make fewer word recognition errors with DW support than the baseline.}
    \end{itemize} 
   
\end{itemize}

\subsection{Methods} 
\subsubsection{Participants.}
We recruited 13 low vision participants \change{(P3 - P15, 9 female and 4 male), whose ages ranged from 30 to 87 ($Mean$ = 64.3, $SD$ = 14.4). Three participants (P5, P8, P15) were legally blind, but still had sufficient functional vision to read visually.} Table  \ref{tab:lv_dem} details participants' visual conditions. No participants had other conditions that cause reading difficulties other than low vision. We recruited participants from a local low vision rehabilitation service. Before a potential participant was recruited, we conducted a screening interview via phone or email to make sure they were eligible for the study. A participant was eligible if they were at least 18 years old and had low vision but still were able to use residual vision to read. All participants completed the study without glasses. Participants were compensated \$20 per hour and were reimbursed for travel expenses.


\subsubsection{Procedure.} \label{sec:proc}
We conducted a single-session study in a well-lit lab. The study lasted 1.5 to 2 hours, including the following phases:

\textbf{\textit{Initial Interview \& Visual Acuity Test.}}
After obtaining participants' consent, we interviewed participants about their demographic information, visual condition, and their challenges with daily reading, as well as their experience with assistive technology.  We then measured their visual acuity using a letter-size ETDRS 1 and ETDRS 2 logMAR chart \cite{ferris1982new}. Participants were instructed to sit at four feet from the eye chart, and for those who could not see the largest row on the chart, we tested their visual acuity at 2 feet. They were asked to read chart 1 with their left eye covered, and then read chart 2 with the right eye covered. We recorded the smallest line that participants can recognize at least three out of five letters correctly. Our visual acuity test covered a range from 20/10 to 20/400. We used reported visual acuity for participants whose visual acuity was outside of the range (Table \ref{tab:lv_dem}).

\textbf{\textit{Gaze Calibration \& Validation.}} We then conducted gaze calibration with participants, including both the 14-dot calibration and line-based correction, as described in Section \ref{sec:cal}. Participants sat in front of a computer screen with an eye tracker. After adjusting their position to achieve a horizontal distance of 65cm to the screen, participants were instructed to sit straight with their back touching the back of the chair to remain that position. We then adjusted screen height to align the participant's eye level with the center of the screen. Participants were asked to keep their body still in the study but small head movement was allowed when necessary. Before calibration, we adjusted the calibration target size for participants until they could locate the center of the target (the black dot) without eye squinting. They then completed the 14-dot calibration. A 5-dot validation followed, which collected the participants' calibrated gaze data when staring at five targets and calculated the accuracy.  Finally, participants completed the line-based correction along with a 4-line validation, where we collected their corrected gaze data when tracking the moving target along four lines and calculated the accuracy. Based on the validation results, we decided which eye's data to use (left, right, or average) following the dominant-eye-based data collection \cite{wang2023understanding}, as well as whether or not to apply the line-based correction.

\textbf{\textit{Tutorial \& Customization.}} After calibration, we conducted a tutorial session to familiarize the participants with GazePrompt and allow them to customize the features. First, we showed participants an example passage and adjusted the font size, font weight, and color for participants so that they could read comfortably without eye squinting. We then introduced the LS and DW support in GazePrompt. For each feature, we demonstrated the two design alternatives and asked them to freely experience the feature on the example passage. During the exploration, participants customized the feature, including the augmentation color for the LS support, and the triggering time threshold for the DW support, until they were fully comfortable and familiar with it. We also asked for participants' feedback and suggestions on each design alternative. Finally, we asked participants to select their preferred design alternative for each feature to use in the following reading tasks. 

\textbf{\textit{Reading Tasks.}} Participants performed multiple trials of \change{reading aloud tasks} in four conditions: (1) without GazePrompt as the baseline, (2) LS support only, (3) DW support only, and (4) both LS and DW support. \change{In all conditions, participant adjusted the reading content to their preferred magnification level and contrast to simulate their daily reading setup. Participants were instructed to read aloud as quickly and accurately as possible without the need of comprehending the content.} We counterbalanced the four conditions using Latin Square \cite{bradley1958complete}. Ten participants read two passages per condition, while the other \change{three (P9, P11, P14) only read one passage per condition due to time constraints}. \change{We randomized the passages across conditions}. We collected participants' gaze data in all reading tasks. 

\change{The passages were selected from CLEAR corpus \cite{crossley2022large}. We first filtered passages with sixth grade level reading difficulty using Flesch Reading Ease \cite{flesch1948new}, and then calculated the cosine similarity between passages based on their Flesch-Kincaid Grade Level \cite{kincaid1975derivation}, Automated Readability Index \cite{kincaid1975derivation}, SMOG \cite{mc1969smog} and word count. As a result, we selected 20 passages with similar difficulty. Eight passages were selected as default passages in our study and the rest were used as back up passages to handle particular situations, such as data loss due to eye tracking failure or system errors.}

\textbf{\textit{Exit Interview.}} We ended our study with a semi-structured interview about participants’ reading experiences with GazePrompt. \change{Participants rated the effectiveness, distraction level, and comfort level of the reading interface when reading in the four conditions on a scale of 1 (strongly negative) to 7 (strongly positive). We also asked for their scores on the perceived accuracy and the learnability of the LS support and DW support, respectively. Participants then discussed their suggestions for improvements and attitudes towards gaze control and traditional manual control.}

\subsection{Analysis}
We collected both quantitative and qualitative data. We first describe our quantitative analysis and then qualitative analysis.


\subsubsection{Effectiveness of LS support}
We first evaluated the effectiveness of the LS support. We defined three measures, including (1) \textbf{line switching time,} that is defined as the time between the last fixation at the end of the prior line and the first fixation that successfully lands on the next line followed by a non-backward saccade along that line. We calculated participants' mean line-switching time across all lines per passage;
(2) \textbf{frequency of line switching deviation}: we define a line switching deviation as an event when a user finishes reading the prior line and intends to switch to the next, their fixations accidentally land on a ``wrong'' line. We counted the total number of such events per passage and calculated the frequency of line switching deviation by dividing the total number of lines in the passage; 
and (3) \textbf{magnitude of line switching deviation}, is defined as the distance between the wrong line and the target line in each deviation. We calculated the mean deviation magnitude per passage. 





We had one within-subject factor \textit{Condition} \change{with two levels---without GazePrompt (\textit{Baseline}), and using LS support only.} 
\change{To investigate the effect of visual conditions, we involved two between subject factors, \textit{VisualAcuity} with two levels---\textit{Low}, \textit{High}---with 20/100 in the better eye as the splitting threshold \cite{wang2023understanding}, and \textit{PeripheralVision} including two levels---\textit{Limited}, \textit{Intact}---based on their self-report visual field}.
To validate the counterbalancing, we involved another within-subject factor \textit{Order}, and found no significant effect of Order on any measures. We checked the normality of each measure using Shapiro-Wilk test. If a measure was normally distributed, \change{we fitted our data with the Linear Mixed-Effects (LME) Model and calculated the ANOVA table to achieve p-values for the fixed effects \cite{kuznetsova2017lmertest};} Tukey's HSD was then used for \textit{post-hoc} comparison if significance \change{was found on interaction between factors}. Otherwise, we used Aligned Rank Transform (ART) ANOVA and ART contrast test for \textit{post-hoc} comparison \cite{elkin2021aligned}. We used partial eta squared ($\eta^2_p$) to calculate effect size, with 0.01, 0.06, 0.14 representing the thresholds of small, medium and large effects \cite{cohen2013statistical}.

\subsubsection{Effectiveness of DW support}
We then evaluated the effectiveness of the DW support.
We define two measures: (1) \textbf{maximum one-pass fixation time on a word}: the maximum time a user fixates on a word within the first pass until their fixation switch to another word; and \change{(2)
\textbf{number of misread words}: the number of words that were read incorrectly by the participants. We identified the misread words by comparing reading content with the audio recordings of participants' reading aloud tasks. Note that we ignored words that are inherently less important and thus often omitted by people when reading, such as articles, prepositions, pronouns and helping verbs, since the errors on these words do not necessarily imply visual difficulty. }

\change{We had one within-subject factor \textit{Condition} with two levels: without GazePrompt (\textit{Baseline}) and using DW support only. Similar to the analysis for LS support, we involved two between-subject factors \textit{VisualAcuity} and \textit{PeripheralVision} to investigate the effect of visual conditions. We also had \textit{Order} as another within-subject factor and found no significant effect of Order on the measure, thus validating the counterbalance. The analysis method mirrors that in prior section. }

\subsubsection{Qualitative Analysis} \label{sec:qual}
We video recorded the interviews and transcribed them using an online automatic transcription service. Our researchers then manually corrected the transcription errors. We analyzed the data using a standard qualitative analysis method \cite{saldana2021coding}. Two researchers independently coded three sample transcripts from three participants using open coding and generated a codebook upon agreement. Each researcher then coded half of the rest interviews based on the codebook, and updated the codebook upon agreement when new code emerged. Finally, we derived themes and categories based on the different aspects (e.g., effectiveness, user preferences) of the evaluated feature.

\subsection{Results}
\subsubsection{Gaze Data Quality}
With our improved calibration procedure that involves line-based correction, the mean angular distance between the estimated gaze position and the target position in the 5-dot validation is \change{0.79\textdegree ~($SD$ = 0.45\textdegree) viewed at 65cm from the screen, which is about 34 pixels on the screen.} This result suggests that participants' gaze data were accurate enough for GazePrompt to function normally, since the smallest font size our participants chose was 48 pixels. Moreover, \change{the percentage of data loss was 5.85\% ($SD$ = 10.1\%) which is much lower than the data loss (about 60\%) reported in prior work \cite{maus2020gaze}. }


\subsubsection{Line-Switching Support}
We report our quantitative and qualitative results about the performance of Line-Switching (LS) Support and participants' preferences on the augmentation design.

\textbf{\textit{Line Switching Time (H1.1).}}
\change{We found a significant effect of \textit{Condition} (LME: $\chi^2(1, N = 26) = 5.99$, $p = 0.014$, $\eta_p^2 = 0.27$) on the line switching time, indicating that the LS support  enabled low vision users to locate the next line faster during line switching.
 We did not see any significant effect of \textit{VisualAcuity} or \textit{PeripheralVision} or their interactions with \textit{Condition} on line switching time. However, we found that among the four participants whose line switching time was not improved (P5, P8, P14, P15), three had peripheral vision loss. This suggests that LS support might not be as effective for people with limited peripheral vision since it could be more difficult for them to notice the augmentations on the next line than people with full peripheral vision. }

\textbf{\textit{Line Switching Accuracy (H1.2).}}
\change{We found no significant effect of \textit{Condition} on the frequency of line switching deviation (ART: $F_{(1,9)} = 0.83$, $p = 0.39$, $\eta^2_p = 0.084$) or the magnitude of line switching deviation (ART: $F_{(1,9)} = 0.19$, $p = 0.67$, $\eta^2_p = 0.021$), indicating no significant improvement of LS support on participants' line switching accuracy. 
This could be partially due to that participants adjusted their reading behaviors to adapt to the line switching difficulty without GazePrompt (see \textit{Scrolling Behaviors Change} section). Additionally, no significant effect of \textit{PeripheralVision} or \textit{VisualAcuity} or any interactions was found on line switching accuracy.
While no significant difference on line switching accuracy, participants generally felt the LS support effectively improved their line switching experiences. As a result, they raised significantly higher scores to the effectiveness of the LS support ($Mean$ = 6.64, $SD$ = 0.50) than magnification without GazePrompt ($Mean$ = 4.82, $SD$ = 1.83, Wilcoxon signed-rank test: $V = 0$, $p = 0.008$).
}

\begin{figure*}[h]
  \includegraphics[width=0.95\textwidth]{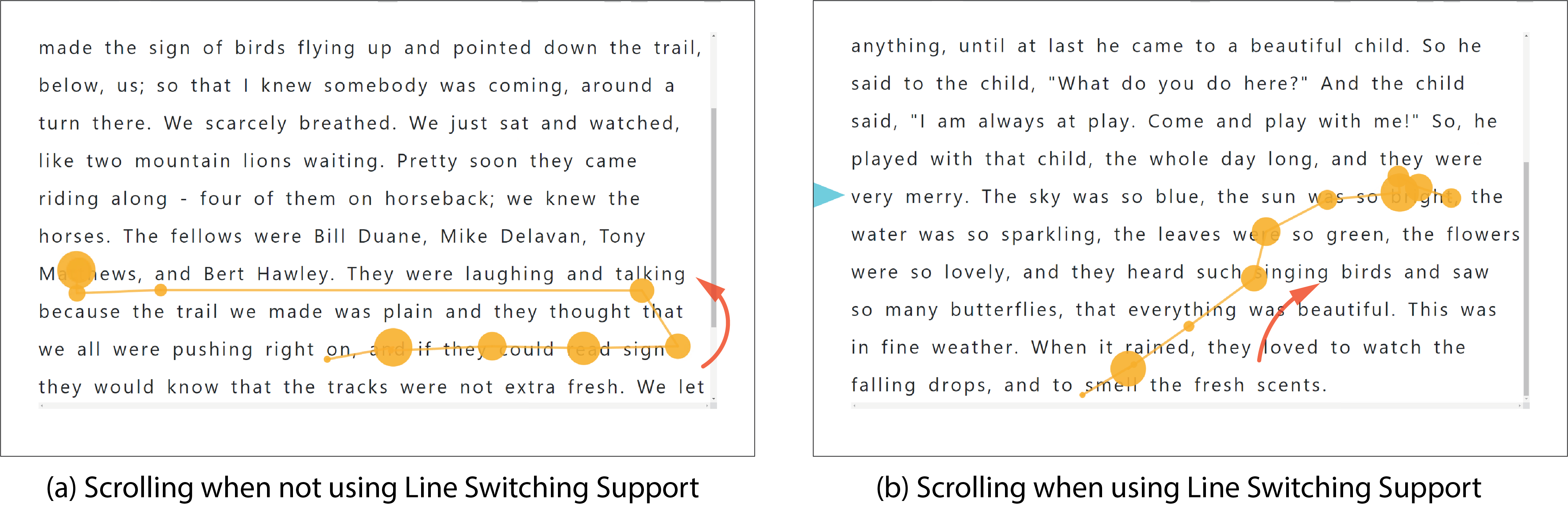}
  \caption{\change{GazePrompt changed participants' scrolling behavior. Orange dots represent fixations and orange line segments represent saccades. (a) P7 scrolled two lines up from bottom while her fixation went up to track the line starting with `Matthews' that she was reading when not using LS Support; (b) P7 scrolled five lines up from bottom while her fixation went up to track the line starting with `very' that she was reading when using LS Support. Red arrow indicates the progressing direction of fixations.}}
  \label{fig:scroll}
  \Description{Figure 5: "This figure shows two passages, one on the left labelled A: Scrolling when not using Line Switching Support; and one on the right labelled B: Scrolling when using Line Switching Support. Both passages have orange circles joined by an orange line with a red curved arrow off to the right side of the circles. The passage on the left shows the circles moving up two lines, but the passage on the right, now including a blue Arrow on the left side of the screen, jumps up five lines from the bottom of the page to the right side of the line indicated by the red arrow."}
    \vspace{-2ex}
\end{figure*}

\textbf{\textit{Scrolling Behavior Change.}}
Since the magnified font required participants to scroll up and down the page to read the whole passage, we further looked into their scrolling behaviors. \change{We found participants scrolled significantly longer distance in a single scrolling event (separated by a pause longer than $100 ms$) when reading with LS support (ART: $F_{(1,12)} = 17.2$, $p = 0.001$, $\eta^2_p = 0.59$), resulting in less scrolling events (Fig. \ref{fig:scroll}b).  When not using GazePrompt, participants usually scrolled only one or two lines at a time to limit the content change, thus easing the line localization during line switching (Fig. \ref{fig:scroll}a). This evidence indicates that the LS support successfully reduces the line localization difficulty, allowing users to scroll more flexibly when reading long paragraphs.}

\begin{figure}[h]
  \includegraphics[width=0.4\textwidth]{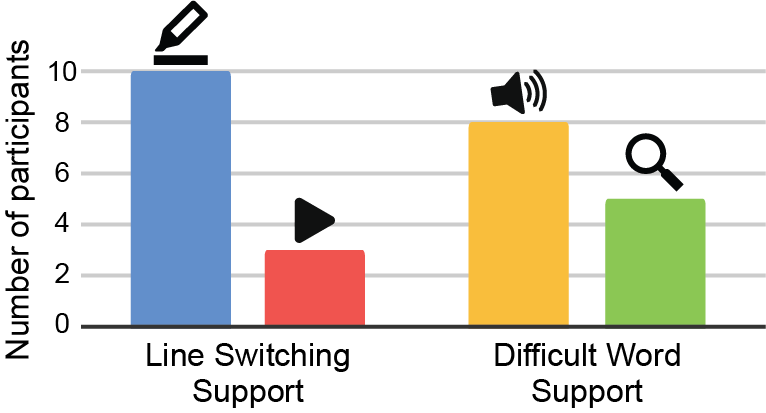}
  \caption{\change{Participants' preference on different augmentation designs of GazePrompt}}
  \label{fig:pref_simple}
  \Description{Figure 6: "This figure is a colored bar chart with the y axis reading "number of participants" from 0 to 10 and the x axis featuring two groups of two bars each, one reading "line switching support" and the other reading "difficult word support". The leftmost bar has a highlight symbol above it and indicates 10 participants preferred highlighting. The second bar has an arrow above it and reads 3 participants. Over the 'difficult word support' text, the left bar shows a speaker icon representing the text-to-speech condition and 8 participants. Finally, the rightmost bar shows a magnifying glass raised to 5 participants."}
    \vspace{-5ex}
\end{figure}

\textbf{\textit{Preference on Augmentation Design.}}
\change{We identified participants' preferences between \textit{Line Highlighting} and \textit{Arrow} design. The majority of participants (10 out of 13) chose Line Highlighting (Fig. \ref{fig:pref_simple}). All of them agreed that LS support improved their focus on the line (P3-P6, P8-P13, P15), thus reducing cognitive load (P13) and eye strain (P4, P10). As P13 mentioned, \textit{``I think [LS support] makes reading the text easier. Because I'm not as focused on [with no support], and I'm kind of going up and down the page. It makes me focus on what I'm really reading instead of worrying about the logistics of reading, like `Okay, where am I on the page?'''} Compared with Arrow, most participants liked that the Line Highlighting augments the whole line instead of just the beginning of lines (P3-P6, P8-P11, P15). This is particularly true for participants with peripheral vision loss since they could barely notice the Arrow as they read towards the end of a line (P13, P15). P6 further commented searching for the arrow made her eyes tired. 

Despite the drawbacks of Arrow, three participants (P5, P7, P14) chose this design since it was less invasive and less distracting. P5 liked that the Arrow entered their visual field only when they needed it, \textit{``When it's on the side, I don't even know what's on the side until I go to my next line. And then I'm like, Oh, it automatically takes me there.''} 

Five participants (P4, P5, P13-P15) appreciated the flexibility with color selection in LS support, which allowed them to customize for their visual condition and preferences. All participants chose the color that created high contrast to the text, with seven of them choosing non-default colors in the system. Participants' choices of color are presented in the appendix (Fig. \ref{fig:pref}).  
Some participants desired even more fine-grained color palette to better customize for their visual abilities in different lighting conditions (P15).}

\textbf{\textit{Potential Improvement on Design.}}
\change{Participants brainstormed potential improvements of the line switching support. Instead of augmenting the whole line, P3 would like a word-level augmentation that guides them through each word as they read on a line. P9 suggested using underlining to augment the line such that the whole line is augmented in a less distracting manner. P15 further suggested placing the arrow on both sides of the screen, which could be more useful for participants with limited peripheral vision loss who could not notice the left side of the screen quickly during line switching. Furthermore, two participants (P9, P13) suggested replacing the arrow with other shapes to further improve their reading satisfaction. The sense of command afforded by arrow made reading less pleasant according to P9, \textit{``The arrow is a command, `Go this way.' Whereas a rectangle would just be saying, `you're on this line.'''}}


\subsubsection{Difficult-Word (DW) Support}
We report our quantitative and qualitative results about the performance of Difficult-Word Support and participants' preference on the augmentation design.

\textbf{\textit{Maximum One-Pass Per-word Fixation Time (H2.1).}} 
\change{While participants felt the DW support made reading faster (P6, P8, P10, P11), we found no significant effect of \textit{Condition} (ART: $F_{(1,9)} = 0.034$, $p = 0.86$, $\eta^2_p = 0.004$) on the maximum time participants fixated on a word in the first pass. This could be because most participants (10 out of 13) chose longer fixation duration threshold to trigger the DW support ($T_0^{DW}$: $Mean = 583 ms$, $SD = 58.3 ms$; $T_2^{DW}$: $Mean = 1913 ms$, $SD = 292 ms$) than default to reduce false positives (Fig. \ref{fig:pref}). We also did not see significant effect of \textit{PeripheralVision} or \textit{VisualAcuity} or any interactions on this measure.}

\change{
\textbf{\textit{Number of Misread Words (H2.2).}} 
We further looked into participants' reading errors. We found that, overall fewer words (30) were misread when using DW support than the baseline (39). 80\% of misread words were recognized as words with similar appearance, such as `though' vs. `through', and `interposed' vs. `interrupted', indicating that the errors were probably caused by visual challenges. 
However, we did not find a significant effect of \textit{Condition} on the number of misread words (ART: $F_{(1,9)} = 1.55$, $p = 0.24$, $\eta^2_p = 0.15$) across participants, neither did \textit{PeripheralVision} or \textit{VisualAcuity} or any interactions. 


\textbf{\textit{Feature Triggering.}} We investigated when the DW support was triggered. Participants on average triggered 17.2 word augmentations ($SD$ = 16.5) when reading each passage using DW support. Out of the 30 misread words under DW support, only two triggered the DW support, and they were all augmented by the Word Magnifier design (P6, P11). This suggests that participants using Word Magnifier might still face difficulty recognizing the words visually.

Overall, the DW support was perceived to be accurate ($Mean$ = 5.58, $SD$ = 0.90) and effective ($Mean$ = 5.45, $SD$ = 1.51)
in recognizing difficult words. However, we acknowledge that the DW support was only helpful when participants noticed the difficulty of the words; the feature would not be triggered if they thought that they recognized the word easily and correctly when they actually did not.
}


\textbf{\textit{Preferences on Augmentation Design.}}
\change{
We report participants' preferences on the \textit{Text-to-Speech} and \textit{Word Magnifier} design. Eight participants chose Text-to-Speech (Fig. \ref{fig:pref_simple}), because audio feedback was perceived to be faster (P3) and easier for people who had stronger auditory senses than visual senses (P3, P14). They felt the audio feedback more human-like and pleasant to use (P8). As P8 explained, \textit{``It's more interactive. It's more rehabilitative. It's like a rehab guy sitting there with me helping me [read].''} Compared with Word Magnifier, Text-to-Speech does not require additional eye movement to scan the magnified word (P3, P12, P15), thus reducing visual burden (P13). Despite the merits of Text-to-Speech, some participants felt audio information took longer to process than visual information (P13, P11). For a reading-aloud task, Text-to-Speech was perceived to be distracting and intrusive (P5, P7, P8, P10, P14) because it sometimes overlapped with participants' speech and therefore interrupted reading process. 

Participants who chose Word Magnifier felt the visual augmentation less distracting and easier to ignore when not needed (P4, P7). For participants who had difficulty recognizing certain single letters in a word due to limited central vision, 
Word Magnifier could be a faster remedy than hearing the whole word. Four out of five participants who chose Word Magnifier had central vision loss, and the other one had intact visual field. P11 described how Word Magnifier helped her recognize words more quickly than Text-to-Speech, \textit{``Because the trouble I was having with the word was generally... just one or two letters... very often it was the first letter of the word. And it popped at me right away [with Word Magnifier]. So then that told me what the word was.''} Furthermore, instead of directly feeding the word to participants via speech, Word Magnifier facilitated sense of agency, making users feel more independent (P11, P12). The drawbacks of Word Magnifier were mostly about the visual design. The magnification level was too high and was not adjustable (P3, P8, P10, P13), and the border was too close to the magnified letters, reducing legibility (P13, P14). 
Moreover, the magnified word could block previous text the user wanted to revisit (P9, P10). Four participants complained that it was easy to lose track on their previous reading position when returning from the magnifier (P5, P8, P11, P12). Two participants with limited peripheral vision also found it difficult to locate the magnifier since its position got outside of their visual field (P13, P15). In fact, no participants with limited peripheral vision chose Word Magnifier.
}


\textbf{\textit{Potential Improvement.}} 
\change{For Text-to-Speech, some participants suggested customization for the voice to make it sound more natural (P5, P8). For Word Magnifier, participants proposed adjustable magnification level (P3), and fixed location (e.g., at the bottom right corner of the screen) to reduce distraction (P5). P3 and P13 also suggested considering different background colors and font for Word Magnifier to make contrast to the original text. P3 also 
suggested augmenting several adjacent words at a time to compensate for potential difficult word identification errors.
}

\subsubsection{Overall Reading Experience with GazePrompt.}
Participants all had positive experiences with GazePrompt. They all agreed both features in GazePrompt were not distracting \change{(LS: $Mean$ = 2.45, $SD$ = 1.69; DW: $Mean$ = 3.82, $SD$ = 2.23)} and easy to learn \change{(LS: $Mean$ = 6.85, $SD$ = 0.38; DW: $Mean$ = 6.89, $SD$ = 0.33)}, but \change{eight} participants said they would be more comfortable using GazePrompt by practicing more. \change{Six} participants like the combination (LS+DW) more than individual features because they improved reading experience from different aspects and can even augment each other. \change{For example, P13 believed the combination of both features were the best, since LS support improved her concentration on words, which in turn made DW support more accurate.}

When comparing gaze-aware augmentations in GazePrompt with manually controlled augmentations (e.g., using a keyboard to switch highlighting to the next line \cite{gowases2011text}), 
nine participants preferred the gaze-aware method \change{(P4, P6-P11, P13, P15) since it was faster (P7, P9, P13, P15), more natural (P13), and more accurate (P11). Due to participants' vision loss, manual control could be cognitively and visually taxing (P4, P5, P13, P11, P15). 
Interestingly, P8 found gaze control rehabilitative since it improved his self-awareness of how he used his eyes. According to P8, \textit{``
[GazePrompt] kind of reminds me that I'm wandering, and my brain wants to correct it. So I'm really envisioning this being a very great rehabilitative tool to read.''} Furthermore, some participants 
suggested combining manual and gaze control to enable more accurate control (P10, P14). For example, P10 would like the Text-to-Speech in DW support to be triggered at a fixated word only after pressing a button. 
However, two participants preferred manual control because they were more familiar with manual control (P3) and the reading distance required by eye tracker made them feel uncomfortable (P12). Participants also brought up the potential mismatch between gaze and their mental status (P3, P9, P12, P13). As P3 explain, \textit{``So my eyes are on something but I [might be] processing something else that I've just read.''}
} 

\section{Study 2: Silent Reading}
\label{study2}
\change{While the well-controlled reading aloud study (Section \ref{study1}) allowed us  to quantitatively examine the effectiveness of GazePrompt, it did not reflect the real-world reading scenario where people usually read silently and focus on content comprehension. To compensate for Study I, we conducted another silent reading study to understand low vision users' experiences with GazePrompt in a more realistic reading context.}

\subsection{Method}
\change{We recruited 13 low vision participants (P13-P25 in Table \ref{tab:lv_dem}), including ten female and three male whose ages ranged from 30 to 85 ($Mean = 58.6$, $SD = 19.6$). Five participants were legally blind but  had functional vision to read. The recruitment method and compensation was the same as Study I (Section \ref{study1}). All participants completed the study, except for P16 who only briefly experienced the features and provided quick feedback due to extensively long calibration.} 

The study lasted 1.5 to 2 hours. \change{The procedure was the same as Study 1 (Section \ref{sec:proc}), except that the reading tasks were silent reading, where participants were instructed to read as quickly as possible in the condition that they could build sufficient understanding of the passage. We asked two simple questions after each passage to ensure participants' basic comprehension on the reading content. 
In the exit interview, in addition to the same set of questions in Study 1, we further prompted for use cases where GazePrompt could be useful.}

We selected reading passages from MCTest MC500 \cite{richardson2013mctest}, a dataset of passages with multiple-choice reading comprehension questions intended for machine comprehension. The passages were fictional stories, reducing the impact of participants' prior knowledge on passage comprehension. We selected 28 passages that were in similar length (about 177-199 words per passage) and at similar difficulty level (about 6th-grade level) according to the Flesch Reading Ease (FRE) score \cite{flesch1979write}. We manually checked each passage, ensuring no inappropriate content was involved. Readings were randomly selected from the 28 passages for each participant. 

\change{As silent reading may involve some complex and even unexplainable gaze behaviors (e.g., fixating on a word while mentally processing previous sentences, or revisiting previous content), our analysis focused only on understanding users' experiences qualitatively. The qualitative analysis method mirrored Study I (Section \ref{sec:qual}).}

\subsection{Findings}
\change{While most findings in this study echoed Study I, we identified insights that were unique to the silent reading tasks, including LS support enhancing comprehension and DW support being used as a confirmation tool. We elaborate these unique findings below.}


\subsubsection{Line-Switching Support for Comprehension} 
Participants' opinions differed on whether the LS support facilitated improved comprehension. Some felt their reading comprehension improved because they could focus on the text better with GazePrompt (P13, P16, P21). One participant felt the background color change caused by Line Highlighting during line switching distracted them, which negatively affected their reading comprehension (P22). 
When asked about the reading scenarios where LS support can be helpful, participants believed it could be particularly favorable when reading long and technical passages that require a certain level of concentration \change{(P13, P16, P23). Two participants mentioned that LS support could be particularly useful in a low lighting condition where their eyes could get tired more easily (P19, P22).} Moreover, P15 indicated that the LS support can be more helpful when reading text with smaller line spacing.  \change{Other than the potential improvements reflected in Study I, P25 further suggested making the Arrow design blink to help him locate the line faster due to his tunnel vision (i.e., severe peripheral vision loss).}

\subsubsection{Difficult-Word Support for Confirmation} 
\change{
Similar to LS support, some participants believed DW support had the potential to aid comprehension (P13, P19, P21). As explained by P21, \textit{``[The Text-to-Speech] helps me just to be able to focus on what I'm reading, and comprehend it without worrying like, `Oh, I'm straining my eyes just by gazing all the time.'''}
Interestingly, besides recognizing difficult words, four participants (P16, P21, P22, P24) mentioned using the DW support as a confirmation tool to the words that they could recognize but were not confident about. As P16 commented, \textit{``when it (text-to-speech) would say it, my brain went `Yes! Yes! That is the word I just saw.'''} P13 even intentionally chose longer fixation duration thresholds to use her gaze as an explicit input (rather than implicit estimation) to trigger the DW support.
Similar to LS support, participants would like to use DW support when reading technical passages (P14, P17, P24), especially those that involve long and unfamiliar words and names (P13-P18, P20, P23).}

\section{Discussion}
We contribute the first research that explored the design space for gaze-aware augmentations to assist low vision people with reading tasks. We built \textit{GazePrompt} to support two important tasks in daily reading: line switching and word recognition. Through a \change{well-controlled reading-aloud} study (Section\ref{study1}) with 13 low vision participants, we demonstrated the effectiveness of GazePrompt. \change{We found that
GazePrompt significantly reduced participants' line switching time (H1.1). While there was no significant improvement on line switching accuracy (H1.2), GazePrompt enabled more flexible scrolling behaviors, not requiring low vision users to scroll only one or two lines at a time to reduce line switching difficulty.} 
For difficult word support, although GazePrompt did not significantly lower the upper bound of participants' word recognition time \change{(H2.1) nor significantly reduce the number of misread words, the overall misread words were reduced across all participants from 39 to 30 (H2.2)}. 
\change{A follow-up silent-reading study with another 13 low vision participants further highlighted the impact and potential of GazePrompt in a more realistic free-form reading context, such as enhancing perceived content comprehension, and enabling confirmation for uncertain words. 
As opposed to manually controlled augmentations, participants expressed strong preference towards gaze-aware augmentations due to their low effort and rehabilitative potential.} 
In this section, we discuss the challenges we encountered in our technology design and implementation, as well as deriving design implications for future gaze-aware technology for low vision. 

\change{\subsection{Eye Tracking for Low Vision}
With improved gaze calibration strategies inspired by prior literature \cite{wang2023understanding} and further vertical drift correction (or line correction), we achieved stable and accurate eye tracking for low vision participants as reflected from both studies. We found line correction can be particularly useful in reading tasks for people with inconsistent dot- and line-validation results. In Study I, one participant (P12) was observed to have low dot-validation error (0.90\textdegree or 39 px), yet high line-validation error (vertical offset: 110 px). This could be due to eye recognition issues with specific head postures, or the fact that the user had inconsistent gaze behavior when performing different tasks (fixation vs. target tracking) to accommodate their visual condition (e.g., central vision loss).
Since the line-calibration mimics the eye movements of reading, we were able to address this inconsistency by applying line offset correction, thus improving the eye tracking accuracy during reading tasks. 

Despite our improvement on the gaze calibration, participants still faced challenges with inaccessible calibration procedure. For example, P1 (who had central vision loss) reported she developed multiple preferred retinal locus (PRL), which hindered current calibration algorithms from learning the correct mapping between the recognized pupil position and where they actually see. P25 had difficulty locating the targets on the screen during calibration due to his severe peripheral vision loss. As such, eight participants across the two studies reported occasional eye tracking issues when using GazePrompt. More research should be devoted to making eye tracking more accessible to people with diverse visual abilities.

In addition to gaze calibration, some participants complained about the distance requirement which made reading uncomfortable. During reading, participants tended to lean closer towards the screen \cite{szpiro2016people, wang2023understanding} and slightly squint even when instructed not to. Such behavior could invariably affect eye tracking accuracy and increase the chance of data loss. Both our work and prior work \cite{wang2023understanding} suggest that future gaze-based assistive technology should consider adopting wearable eye trackers and communicate eye tracking issues promptly to the user when necessary. 
}

\change{\subsection{Design Implications}}
Inspired by participants' preferences on different augmentations as well as usability issues encountered using GazePrompt, we discuss potential design implications for future gaze-aware low vision aids.

\textit{\textbf{Intelligent Line Identification Mechanism.}}
Our line identification algorithm could successfully locate the line the user intended to read in most cases when the user revisits or jumps to a new line. However, such design could be less ideal for users with severe vision loss who requires excessively long time searching for lines during reading. For example, line switching is difficult for people with severe visual field loss (e.g., \change{P25}). Because of their limited visual field, they need to back trace on the line very slowly to make sure they are on the right track. However, with our line identification mechanism, this slow search behavior will be misidentified as a ``line jumping'' behavior with the augmentation moving to the wrong line. As such, the Line-Switching augmentation will be less usable. Prior work has used gaze data to predict users' reading comprehension \cite{ahn2020towards} and their intention in social scenarios \cite{park2013predicting}. No prior work had used gaze data to predict fine-grained fundamental gaze behaviors in reading, not to mention for low vision users. In light of this issue, future researchers should consider building a gaze behavior dataset that covers people with diverse low vision conditions, which would support the development of more intelligent and personalized algorithms to recognize low vision users' gaze behavior in reading. 

\textit{\textbf{Customization and Adaptation to Users' Reading Habits.}}
GazePrompt provided a certain level of customization for visual augmentations. Participants could adjust the color for Line-Switching support and the fixation duration threshold for Difficult-Word support, which are two important parameters that made GazePrompt useable for low vision participants. However, participants would like additional customization options. For example, they would like augmentations to be adapted to their reading habits, such as reading speed (e.g, P8). Even with our experimental fixation duration adjustment, some participants still experienced under- or over-sensitivity that increased the number of false positives. Since low vision people can experience diverse visual conditions, it is important to provide sufficient and fine-grained customization to maximize the efficacy of gaze-aware augmentations designed for the low vision population. Moreover, in the field of HCI, adaptive user interfaces have been studied to deliver smooth and convivial user experiences \cite{alvarez2009current, hussain2018model, gajos2008predictability}. Similar ideas can be applied to adapt system parameters to each low vision user's unique reading habits.

\textit{\textbf{Gaze-Aware Technology Requires High Gaze Control Ability.}}
While most participants were optimistic about the gaze-aware augmentations, several pointed out the potential issues low vision users have when using gaze as an input modality. GazePrompt, as a system that is fully controlled by users' gaze, provided high responsiveness, enabling more targeted assistance in dynamic reading processes. However, the requirement of using eyes as input for a long time can cause eye strain (e.g., P21). Moreover, some low vision users who could not effectively control their eye movement due to some visual conditions (e.g., Nystagmus) could not use their eyes as the system requires. \change{Given low vision users' difficulty in hand-eye coordination, and the strain caused by using gaze as the sole approach for manipulation, integration of gaze control and manual control should be considered to overcome the issues with either interaction modality. For example, gaze can be used for selection, and manual control for simple manipulation (e.g., tapping, pressing a button) \cite{pfeuffer2016gaze, pfeuffer2014gaze}. As such, convenience and accuracy of eye tracking can be preserved without involving significant visual or physical stress. Prior work has shown that the combination of gaze control and manual control can improve task completion efficiency without causing significant eye discomfort for sighted participants \cite{pfeuffer2016gaze, pfeuffer2023palmgazer}. Future research should comprehensively investigate low vision users' experience with gaze and manual control to derive new interaction paradigms for low vision users. Moreover, since low vision users manifest a wide range of visual abilities, the involvement of gaze control should be customizable.}



\subsection{Limitations}
Our research has limitations. \change{Although we involved 23 participants in the two studies for system evaluation, we only had 13 participants in each study given the difficulty of recruiting low vision people (e.g., limited mobility). Therefore, the power of our statistical analysis result is limited. Future work is needed to fully examine the potential of gaze-aware reading aid with sufficient number of participants representing different visual conditions. Second, we did not compare the effectiveness of gaze-aware reading augmentations with manually-controlled counterparts during evaluation. Therefore, participants' response about manually controlled reading augmentations were all based on their prior experience with conventional input method (e.g., keyboard and mouse ), which might not reflect their true preference. Future work should compare gaze-aware low vision aids with manually controlled aid to draw conclusions about the control modality more objectively.}

\section{Conclusion}
In this paper, we presented GazePrompt, a reading aid system that provides gaze-aware augmentation for low vision users. Our user studies with \change{23 participants showed that GazePrompt improved participants' line-switching performance and difficult word recognition performance, and could potentially enhance comprehension. Participants discussed their preferences on the augmentation design of GazePrompt and their attitudes towards gaze-aware versus manually controlled reading augmentations. Based on our quantitative and qualitative findings, we discussed eye tracking challenges for low vision users and derived design implications for future gaze-aware low vision aids.}

\begin{acks}
This work was partially supported by the University of Wisconsin—Madison Office of the Vice Chancellor for Research and Graduate Education with funding from the Wisconsin Alumni Research Foundation.
\end{acks}

\bibliographystyle{ACM-Reference-Format}
\bibliography{sample-base}

\appendix
\renewcommand\thefigure{\thesection.\arabic{figure}}    
\setcounter{figure}{0}  
\renewcommand\thetable{\thesection.\arabic{table}}    
\setcounter{table}{0}  
\section{Appendix}

\begin{figure*}[h]
  \includegraphics[width=0.95\textwidth]{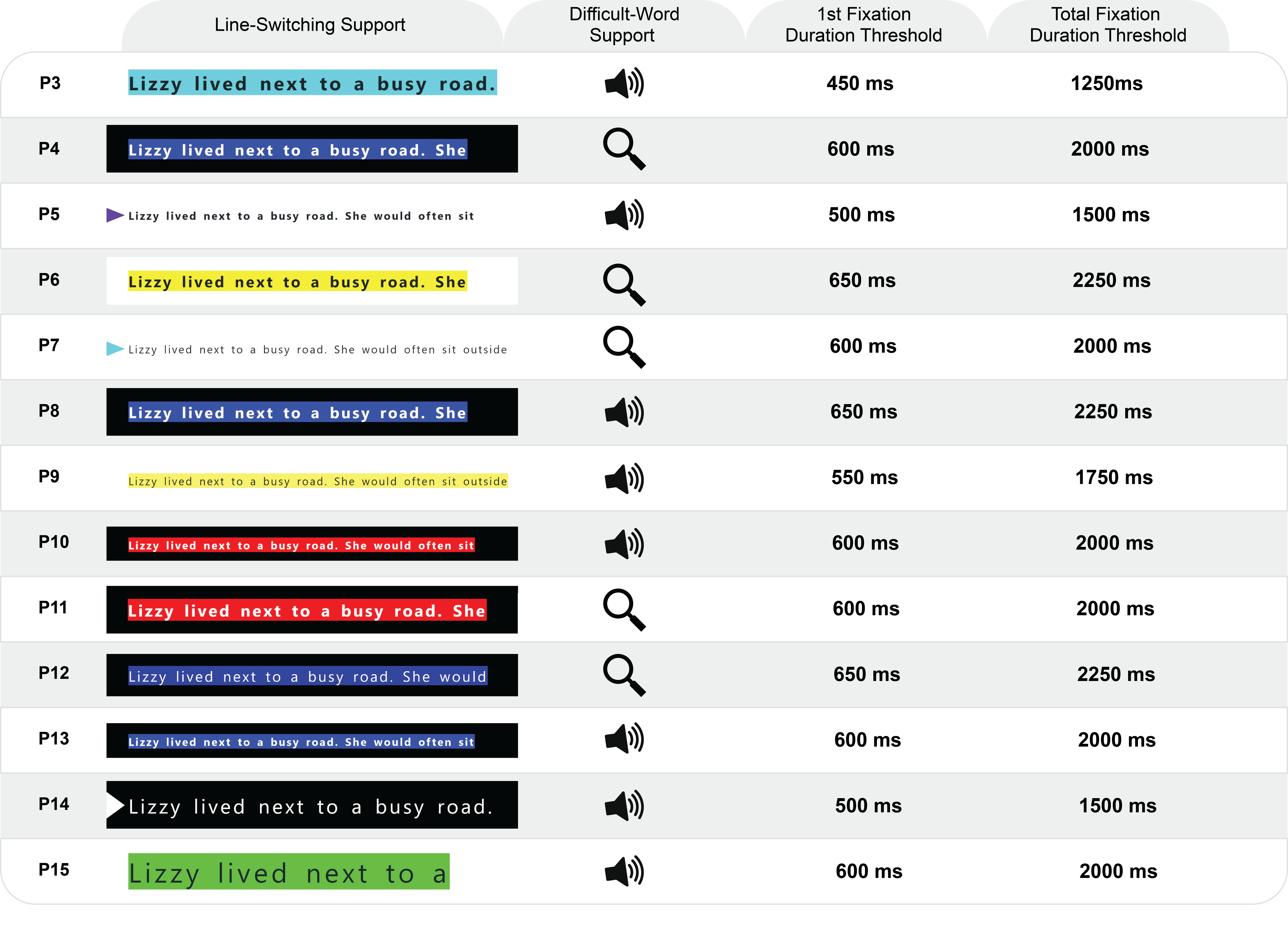}
  \caption{Participants' preference on the two features of GazePrompt. The first column shows participants' preference on Line-Switching support design along with their text setting. The second column shows participants' preference on Difficult-Word support, where the magnifying glass icon represents Word Magnifier, and the speaker icon represents Text-to-Speech.}
  \label{fig:pref}
  \Description{Figure A1 in Appendix: "This figure shows a table with the following columns:
1. participant IDs;
2. an illustration of each participants' preferred line switching support method (like "blue highlighting on white text on a black background" or "dark blue arrow next to bold black text on a white background");
3. the participants' preferred difficult word support method (an icon of either a speaker to represent text-to-speech or a magnifying glass to represent the magnifier);
4. the participants' first fixation duration threshold, in milliseconds, the shortest being 450ms and longest being 650ms; and
5. the participants total fixation duration threshold, ranging from 1250ms to 2250ms."}
    \vspace{-2ex}
\end{figure*}

\end{document}